\documentclass[draft]{agujournal2019}
\usepackage{subcaption}
\usepackage{comment}
\usepackage{lineno}
\usepackage{soul}

\newcounter{SuppCount}
\newcommand{\SuppRef}[1]{%
	\stepcounter{SuppCount}%
	#1\theSuppCount
}

\draftfalse

\journalname{Geophysical Research Letters}

\begin{document}

\title{Homogeneous soil moisture fields suppress Sahelian MCS frequency}
\authors{Ben Maybee\affil{1}, Cornelia Klein\affil{2}, Christopher~M. Taylor\affil{2,3}, Helen Burns\affil{1}, and John~H. Marsham\affil{1}}

\affiliation{1}{School of Earth and Environment, University of Leeds, Leeds, UK}
\affiliation{2}{UK Centre for Ecology and Hydrology (UKCEH), Wallingford, UK}
\affiliation{3}{National Centre for Earth Observation (NCEO), Wallingford, UK}
\correspondingauthor{Ben Maybee}{b.w.maybee@leeds.ac.uk}

\begin{keypoints}
\item Suppressing sub--synoptic soil moisture heterogeneity ($<$1000km) decreases peak MCS population by 23\%, without weakening convective intensity.
\item Mesoscale dry soil patches (100--500km) modify the Sahelian boundary layer such that conditions are more favourable for mature MCS passage.
\item 
Increased insolation in cloud--free slots can replace dry SM as a source of favourable MCS conditions
, in tandem with background monsoon flow.

\end{keypoints}

\begin{abstract}
    Understanding controls on Mesoscale Convective Systems (MCSs) is critical for predicting rainfall extremes across scales. Spatial variability of soil moisture (SM) presents such a control, with $\sim$200km dry patches in the Sahel observed to intensify mature MCSs. Here we test MCS sensitivity to spatial scales of surface heterogeneity using a framework of 
    {78 Unified Model} experiments initialised from scale--filtered SM. We demonstrate the control of SM heterogeneity on MCS populations, and the mechanistic chain via which spatial variability propagates through surface fluxes to convective boundary layer development and storm environments. When all sub--synoptic SM variability is homogenised, peak MCS counts drop by 23\%, whereas maintaining small--scale variability 
    maintains primary initiation rates, reducing the drop in MCS totals
    . In sensitivity experiments, boundary layer development prior to MCSs is similar to that over mesoscale dry SM anomalies, but 
    driven by 
    cloud--free slots of increased shortwave radiation. This reduces storm numbers and potential predictability.
\end{abstract}

\section*{Plain Language Summary}

Tropical rainfall is dominated by Mesoscale Convective Systems (MCSs), large, long--lived organised clusters of thunderstorms. Storm properties are determined in part by the balance of surface fluxes of heat and moisture. Over semi--arid land regions such as the African Sahel, these fluxes are controlled by soil moisture (SM) content. Observations show that the impact of SM on MCS processes then varies depending on the scale of spatial variability of the SM field. Here we assess the sensitivity of Sahelian MCSs to scales of SM variability by running 
78 
high--resolution atmospheric simulations in which only the early morning land surface is altered. In one experiment set we suppress all SM variability below $\sim$1000km, yielding a homogeneous land--state, while another experiment reintroduces small--scale SM variability. Both experiments yield a significant reduction in MCS numbers, with the strongest effect in experiments with homogeneous SM. Warm planetary boundary layer conditions which support subsequent MCS activity are typically found over mesoscale dry SM anomalies. When SM is homogenised, high insolation plays an increasing role in creating the locally warm conditions favouring convection. This reduces storm numbers, and likely also predictability, since cloud--free areas are more diffuse and transient than SM patches.

\section{Introduction}

Mesoscale Convective Systems (MCSs), large organised convective storms which play a prominent role in tropical weather and climate, show significantly different characteristics over land versus oceans (e.g. \citeA{Abbott2025land}). This high--level comparison highlights the striking fact that the balance of surface turbulent fluxes exerts a major influence on one of the primary sources of tropical rainfall. Deep convective organisation and intensity is determined by atmospheric drivers such as moisture, Convective Available Potential Energy (CAPE) and wind shear, the local balance of which characterises different MCS hotspots and explains their geographical occurrence across spatial scales \cite{Laing2000large,Schumacher2020formation}. The inherent orographic, biophysical and hydrological heterogeneity of the land--surface leads to spatial variability in latent and sensible heat flux partitioning, and thereby planetary boundary layer (PBL) development \cite{Carson1973development,Betts1995fife,Huang2013impact}, altering local balances of convective drivers.

Where soil moisture (SM) deficits strongly influence the surface flux partition of available energy there can be significant feedbacks on convective rainfall  \cite{Koster2004regions}. In principle, a chain of local couplings to the atmospheric column can affect the initiation and development of convection (LoCo, \citeA{Santanello2018land}), providing analytical understanding of when convective triggering may occur \cite{Findell2003atmospheric,Bhowmick2018analytical}. However, observations suggest that non--local mechanisms dependent on the spatial variability of SM (SM heterogeneity henceforth) must be invoked to fully understand land--atmosphere interactions with precipitation (P; \citeA{Guillod2015reconciling,Hsu2017relation}).

This is especially the case for MCSs, which respond to SM heterogeneity at scales from tens to hundreds of kilometres through the generation of daytime mesoscale circulations \cite{Taylor2007observational}. In the Sahel region of West Africa, the likelihood of MCS initiation is doubled over $\sim$30km SM gradients due to enhanced local convergence \cite{Taylor2011frequency}; the region’s relatively uniform topography and absence of irrigation isolate this signal, but such controls on initiation are observed 
in other continents 
\cite{Taylor2015detecting,Teramura2019observed,Gaal2021soil}. At the synoptic scale, \citeA{Barton2025storm} show that meridional SM gradients intensify mature MCSs by strengthening vertical wind shear in multiple semi--arid regions, including West Africa. Here regional SM strongly influences the midlevel African Easterly Jet (AEJ), an integral feature of the West African Monsoon (WAM) circulation \cite{Cook1999generation,Talib2022sensitivity}.

The Sahel is characterised by a negative SM--P feedback \cite{Taylor2013modeling}. In particular, mesoscale ($\sim$200km) dry SM anomalies are observed to intensify mature MCSs \cite{Klein2020dry}. Filling the spectrum of spatial heterogeneity, such patches are found to build favourable conditions for intense convection through a combination of convergence, increased instability and wind shear. This feedback imbues the Sahelian land--surface with short--term predictive skill \cite{Taylor2022nowcasting}, with mesoscale wet patches deposited by storm rainfall exerting multi--day suppression of following events \cite{Taylor2024multiday} and causing severe wet--bulb temperature events \cite{Chagnaud2025wet}. MCS activity over the Sahel is intensifying under climate change \cite{Taylor2017frequency}, with potentially fewer, more extreme storms in the future \cite{Kendon2019enhanced}. This could enhance SM heterogeneity at the mesoscale (e.g. \citeA{Hsu2025robust,Guilloteau2025amplified}) and thereby likely further exacerbate future extremes.

Such feedbacks are poorly represented in coarse--gridded global climate simulations that must parameterise deep convection \cite{Taylor2012afternoon}. The sign of the SM--P feedback depends on the treatment of convection \cite{Hohenegger2009soil}, especially regarding the impact of heterogeneity \cite{Taylor2013modeling}, while \citeA{Lee2024weaker} found a convection permitting (CP) global simulation weakens positive temporal SM--P correlations. Kilometre--scale CP models provide significantly improved representation of both land--surface features and MCS dynamics, realistically simulating the diurnal cycles of MCSs \cite{Prein2015review} and their responses to environmental drivers \cite{Maybee2024wind}. However, MCS populations in global CP models show other systematic biases \cite{Feng2023mesoscale,Feng2025mesoscale}, making process--analysis of MCS drivers, including land--atmosphere interactions, an integral component of model evaluation.

It is therefore important to interrogate the sensitivity of MCS populations to changes in SM heterogeneity across scales, since such heterogeneity is itself resolved in kilometre--scale models. We explore this question in the Sahel using two sets of CP sensitivity experiments 
comprising daily restarts from an evolving SM field with filtered variability at selected scales. Other 
studies tackle small--scale effects in other regions \cite{Gaal2024identifying,Paccini2025influence}, however our unique setup enables an isolation of mesoscale feedbacks 
{on mature systems}. We show that suppression of SM heterogeneity significantly diminishes MCS numbers, and elucidate the land--atmosphere interactions responsible. 





\section{Scale sensitivity experiments}

We utilise high--resolution CP model experiments run with  the Met Office Unified Model (MetUM; \citeA{Brown2012unified}). All experiments feature a 1.5km grid spacing; the RAL3.1 physics configuration \cite{Bush2025third} 
of default regional modelling settings and parameterisations 
of sub--grid processes; and the JULES land--surface scheme \cite{Best2011joint}.

A simulation spanning 25/07/2006 to 09/03/2006 was initialised at 00UTC on 23/07/2006 using ERA5 atmospheric conditions \cite{Hersbach2020era5} and (4 level) SM derived from a previous 10 year CP MetUM experiment \cite{Stratton2018pan}. Since our simulation used identical soil properties as this 
model
, the SM profile was already spun--up and we analysed all but the first 48 hours. The 1.5km West Africa domain is nested within a 4km simulation (Fig~\SuppRef{S}a) that itself takes ERA5 boundary conditions every 6 hours. From the 1.5km simulation we extract the 06UTC atmospheric and land--surface state from 25/07/2006 to 01/09/
2006
. These fields then provide the initial conditions for 39 groups of restarted 48 hour long runs where the only alteration made to the initial conditions is in the top (10cm deep) SM layer. 

\subsection{Control and SM(LargeOnly)}

Our first target is to suppress all SM heterogeneity while maintaining large--scale patterns that feed--back on synoptic atmospheric features such as the AEJ \cite{Cook1999generation}. We apply a Gaussian filter with kernel radius of 600km to the 06 UTC SM field --- this suppresses all sub--synoptic scale variability while only marginally weakening the meridional gradient across our primary study region, the Sahel (Figs.~1a,b; -12$^\circ$---18$^\circ$E, 9$^\circ$---19$^\circ$N).

We also reinitialise companion 48--hour Control runs, in which no changes are made to the 06UTC land surface. All anomalies are calculated against Control means throughout. Comparing the time evolution of the simulations' respective power spectra (Fig.~S1b) shows the expected suppression of variability in SM(LargeOnly) for scales below 1000km throughout the full experiment, but with a modest recovery in sub--200km variability due to convective activity reintroducing SM features at these scales. Figure~\SuppRef{S} shows the mean state 
of the WAM system remains very similar between SM(LargeOnly) and Control; any differences between the models can therefore be attributed to the impact of the land--surface on sub--synoptic scale processes.

\subsection{Wavelet reconstruction and SM(Large+Small)}

The SM(LargeOnly) experiments remove SM variability $<$1000km. In order to 
isolate 
the control of 
mesoscale SM variability on 
mature convection
, we also run a SM(Large+Small) experiment where we reintroduce SM patchiness at scales $<$100km, excluding only the mesoscale 100--500km scale range. We quantify variability at a given spatial scale using a wavelet transform onto an isotropic Marr/Mexican--hat basis initialised across 40 scales starting from 4km (see \citeA{Wang2010two,Klein2018wavelet}). To construct a 06UTC SM field from which mesoscale variability is removed, we set the wavelet coefficients for scales above 100km to zero and apply an inverse transform on the same basis. This field, which comprises a reconstruction of all small--scale ($<$100km) variability, is then added to the Gaussian--smoothed field used to initialise SM(Large Only). The 39 resulting SM fields are used to initialise the second experiment set, SM(Large+Small). 

The reinstatement of small spatial scales in SM(Large+Small) versus SM(LargeOnly) is clearly visible from Fig.~1c, while intermediate scales are still missing versus Control. Analysis of power spectra confirms this (Fig.~S1c), with variability suppressed versus Control at the mesoscale. 
{Synoptic mean states are again comparable (Fig.~S2).}


\section{SM derived changes to PBL}
\label{sec:patch_changes}

\begin{figure}[t]
    \centering
    \includegraphics[width=\textwidth]{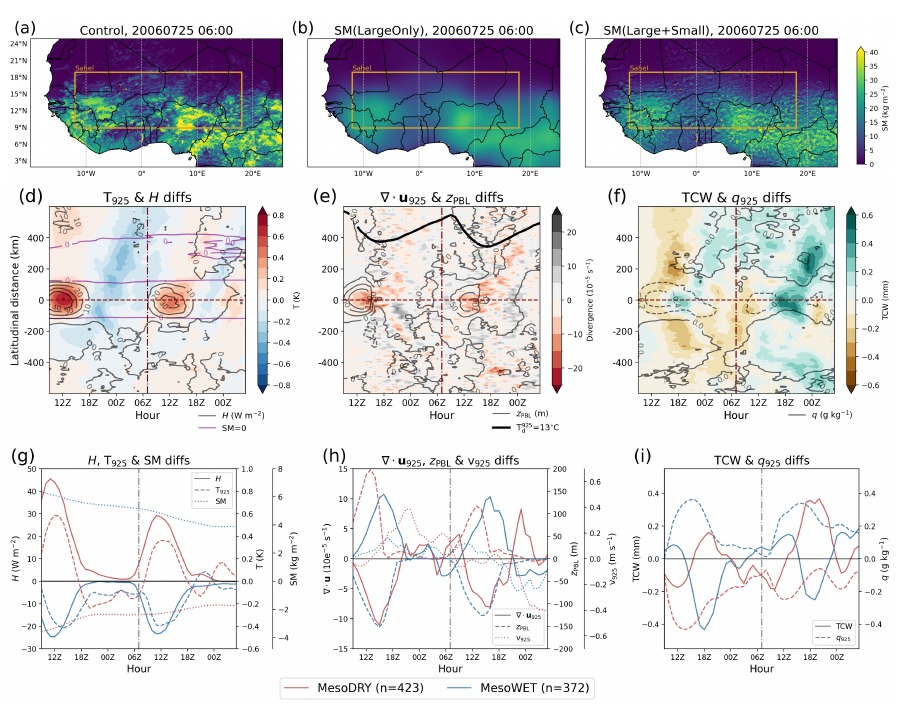}
    \caption{\textit{(a-c)} Representative 06UTC surface-layer SM content for each simulation; boxes show Sahel domain. \textit{(d-f)} Composite evolution about MesoDRY locations of difference between Control and SM(Large+Small) \textit{(d)} sensible heat flux and 925hPa temperature; \textit{(e)} PBL height and 925hPa divergence; and \textit{(f)} TCW and 925hPa humidity. Control mean ITD (bold 925hPa 13$^\circ$C dewpoint contour) shown in \textit{(e)}. \textit{(g-i)} Area--averaged evolution of same field differences about MesoDRY and MesoWET locations, also with \textit{(g)} SM and \textit{(h)} 925hPa meridional wind. All averages taken over 150km slices.
    }
\end{figure}

Before considering MCSs, we first isolate mesoscale SM patches' impact on PBL development without reference to convection. To do so, we identify locations where mesoscale patches were removed by wavelet--transforming 
each D1 
09UTC 
{(+3hr)} SM difference field between Control and SM(Large +Small), isolating regions where the resulting 
{variance--normalised} power spectrum is 
{$>$2 (i.e. $\sim\!95\%$ significant)
. Sample locations are then power maxima for scales between 150--650km, with the largest scale used where a region shows multiple maxima. Our sampling ensures we only consider locations where strong mesoscale SM anomalies were removed, resulting in 423 cases for dry (MesoDRY) and 372 cases for wet (MesoWET) anomalies.

 MesoDRY control on PBL development over two diurnal cycles is shown by composite Hovmoellers (Figs.~1d--f; MesoWET counterparts Fig.~\SuppRef{S}) of the difference between Control and SM(Large +Small) fields. Since the only change to our sensitivity experiments is the initial SM, such differences isolate the effect 
 that SM anomalies in Control have on atmospheric fields. Centring on MesoDRY locations, i.e. where there is a dry patch
 , we find a mesoscale region of strong sensible heat fluxes $H$. PBL temperatures $T$ are thus elevated, and there is colocated, mesoscale growth of PBL height ($z_{\rm PBL}$;~1e). This generates low--level convergence ($\nabla\cdot\mathbf{u}$) soon after midday, enhancing southerly meridional winds $v$ (1h) while the intertropical discontinuity (ITD) shifts southward.

Dry patches cause a mesoscale suppression of latent heat flux $L_cE$ (not shown), reducing low--level moisture near the patch (Fig.~1f; vertical sections Fig.~\SuppRef{S}b). There is a complementary relative moistening of air above $z_{\rm PBL}$, caused by increased entrainment of dry air into the PBL under elevated $H$. Overall, total column water (TCW) increases from 14UTC onwards (Fig.~1f,i). This is driven by the mesoscale low--level convergence, which generates net ascent and from 15UTC--onwards, column moisture convergence (Fig.~S4g), with the feedback further enhanced on day 2 (D2). Moreover, the dynamical changes enhance the local monsoon flow: nocturnal north--eastwards propagating bands of enhanced TCW (Fig.~1f), reduced temperatures (1d), increased southerly winds (1h) and meridional moisture flux (Fig.~S4g) indicate a strengthened WAM flow, peaking overnight \cite{Parker2005diurnal}.

Figs.~1g--i show MesoDRY/WET variables averaged across a 150km box centred on identified patches. As expected, MesoWET locations show an opposite effect to MesoDRY, with a $H$ deficit in Control versus SM(Large+Small) and $L_cE$ instead increased, enhancing low--level moisture favourable for convection. However, absolute temperatures are reduced and $z_{\rm PBL}$ is lowered, strengthening divergence above the wet patch and yielding no monsoon enhancement. There is instead strengthened moisture divergence relative to MesoDRY (Fig.~S4h), leading to decreased TCW from mid--afternoon onwards above the wet patch, above which we also find net subsidence. Indirect feedbacks of $H$ on column moisture, mediated by dynamical changes to the WAM, are again stronger than direct feedbacks from surface evaporation, yielding 
enhanced TCW over dry SM patches and suppressed TCW over wet patches.

In summary, mesoscale PBL deepening driven by enhanced $H$ from underlying dry SM anomalies causes enhanced convergence and localised net ascent. In the Sahel, these dynamical feedbacks over dry soils enhance the monsoon flow and increase column moisture.

\section{SM--driven changes to MCSs}

We now take a convection--centred perspective
{, focussing on mature storms}. To identify MCSs we apply the simpleTrack algorithm \cite{Stein2014three}, adopted and described in \citeA{Maybee2025how}, applied to brightness temperatures ($T_b$) calculated from total outgoing longwave radiation. Storms are tracked over the model domain and comprise snapshots where $T_b<241$K over a minimum area of 1000km$^2$, with candidate MCSs reaching a lifetime maximum area of at least 5000km$^2$ and minimum $T_b<$223K. For these tracks, rainfall volumes and extremes are calculated (from 0.1$^\circ$ regridded precipitation), with MCSs required to reach a maximum rainfall rate above 1mm hr$^{-1}$ during the storm's lifetime.

\subsection{Storm characteristics}

\begin{figure}[t]
    \centering
    \includegraphics[width=\textwidth]{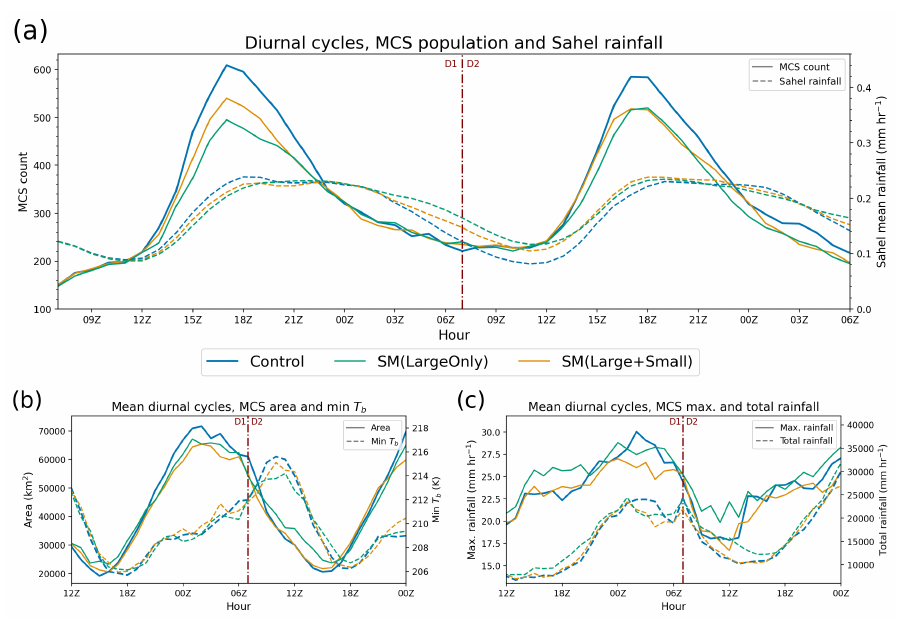}
    \caption{\textit{(a)} Hourly total counts of tracked Sahel MCS snapshots (solid lines) and regional mean rainfall (dashed) across all 39 members of each simulation. \textit{(b)} Mean MCS snapshot areas (solid) and minimum $T_b$ (dashed) between 12 UTC D1 and 00 UTC D2. \textit{(c)} Mean hourly MCS maximum (solid) and total (dashed) rainfall rates over same period.
    }
\end{figure}

Figure~2a shows the combined diurnal cycles of the Sahelian MCS population across each set of simulations. All show identical phasings, with a 17UTC peak in activity following primary initiation. However, amplitudes are significantly altered in the experiments: in SM(LargeOnly) the D1 peak storm count is 23\% lower than in Control. On D2 in SM(LargeOnly) the population begins to recover, at 4\% higher than D1 and 13\% lower than the D2 Control peak. In SM(Large +Small), the drop in storm numbers is smaller than SM(LargeOnly), with the peak 13\% lower than Control, but this does not recover into D2. We find no change in storm speeds and lifetimes in either experiment (not shown).

In both experiments, afternoon D1 mean Sahel rainfall and MCS activity is reduced, with regional totals and storm numbers equivalent to Control overnight. Concurrently, MCSs grow larger in Control than in the experiments (Fig.~2b), but with comparable convective intensities as measured by cloud temperatures and storm rainfall (Figs.~2b,c; \SuppRef{S}a). Fewer primary initiations in SM(LargeOnly) reduces the number of small systems (S5b,c), artificially inflating daytime mean MCS areas and rainfall. From 03 to 18UTC D2 however, regional rainfall totals are lower in Control. This now reflects a relative intensification of convective activity in SM(LargeOnly): there are comparable MCS numbers (Fig.~2a), of all sizes (Fig.~S5c), but with larger, colder anvils (Fig.~2b) and stronger rainfall (2c), coincident with higher mean regional rainfall and cloud cover (Fig.~S2c). Relative recovery in Control then comes during the second afternoon.

We thus conclude that in the Sahel, SM heterogeneity exerts a significant positive feedback on afternoon MCS populations which is maintained through the diurnal peak and into the night. Storms that persist until the early morning of the next day, however, are relatively weakened. 

\subsection{Mechanisms}

\begin{figure}[t]
    \centering
    \includegraphics[width=\textwidth]{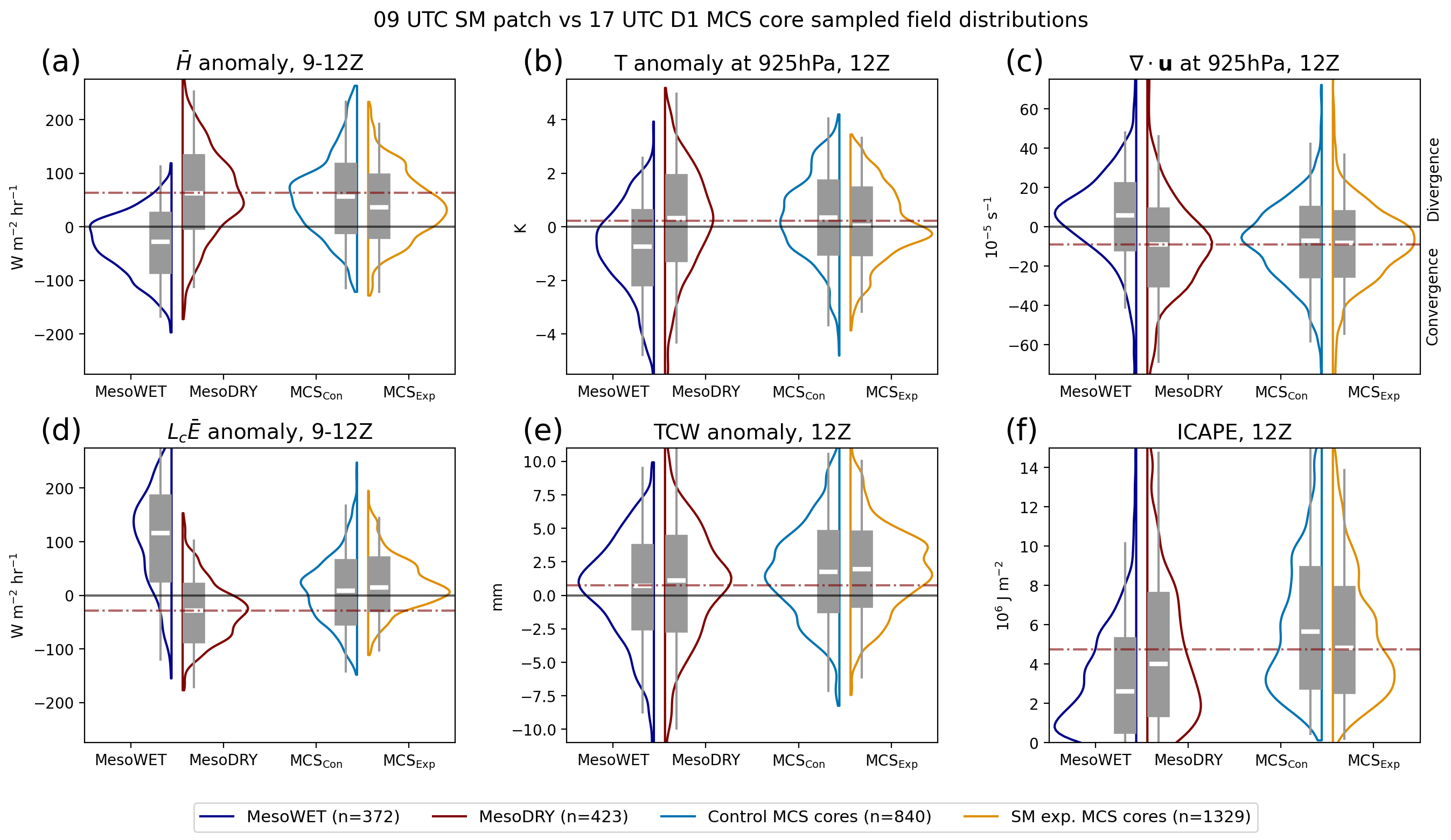}
    \caption{Violin and box plots of 1$^\circ$ mean \textit{(a)} sensible heat flux and \textit{(b)} 925hPa temperature anomalies; \textit{(c)} 925hPa divergence; \textit{(d)} latent heat flux and \textit{(e)} TCW anomalies; and \textit{(f)} integrated CAPE. Distributions centred on 
    {D1} 09UTC MesoDRY/WET and 17UTC MCS core locations, with sensitivity experiment MCS (MCS$_{\rm Exp}$) conditions aggregated from both experiments. All 12UTC fields unless stated otherwise
    {, legend specifies sample sizes}. SM patches sampled in Control, horizontal red lines show MesoDRY means.
    }
\end{figure}

To understand the land--atmosphere interactions which cause these impacts on regional MCSs, we examine land and atmospheric conditions prior to convective cores in mature MCSs at the diurnal peak (17UTC) of convection. We define mature MCSs as $T_b<241$K contiguous cloud covering an area $>$15000km$^2$, within which cores are identified as multi--pixel regions where the 500hPa updraft velocity is greater than the 99.5$^{\rm th}$ percentile of in--storm values. We exclude cores located over orography above 450m; where there was an MCS anvil covering the same location at 15UTC; and where the enveloping MCS initiated less than 150km away \cite{Klein2020dry}. This avoids situations where local pre--core conditions have already been disturbed by the same MCS.

We compare distributions of 
{D1} 12UTC pre--core environmental fields (Fig.~3) between Control (MCS$_{\rm Con}$) and the 
{aggregated} sensitivity experiments (MCS$_{\rm Exp}$). Alongside, we show conditions developed by mesoscale SM patches by sampling Control fields at MesoDRY and MesoWET locations, hereby evaluating which distributions resemble pre--MCS conditions. Consistent with observations \cite{Klein2020dry}, MCS$_{\rm Con}$ cores are associated with pre--convective positive $H$ and low--level temperature anomalies, comparable distributions of which are found over MesoDRY but not MesoWET locations (3a,b). Similar, but weaker, conditions are still evident for experiment cores despite the absence of mesoscale SM anomalies. This suggests the importance of an alternative mechanism establishing these environments in the sensitivity experiments. Meanwhile the low--level convergence generated by MesoDRY SM perturbations resembles MCS distributions (3c), indicating they provide favourable environmental conditions for mature MCSs.

Latent fluxes are not a key influence on mature MCSs even in sensitivity experiments (3d), with high $L_cE$ anomalies about MesoWET patches manifestly not favoured by storm cores. However, cores remain located at positive low--level humidity (Fig.~\SuppRef{S}a) and TCW (Fig.~3e) anomalies, emphasising the background monsoon flow's role in generating favourable thermodynamic environments. As anticipated from Sec.~3, distributions of TCW about MesoDRY patches are therefore closer to those for MCSs than MesoWET, despite stronger low--level moisture anomalies about the latter.

Analogously to TCW, we can measure total column instability through integrated CAPE,
\begin{equation}
	\textrm{ICAPE} = \frac1{g}\!\int\!dp\,\textrm{CAPE}(p)\,,
\end{equation}
where CAPE$(p)>$0. 
Using soundings calculated from 1$^\circ$ mean profiles
, we find the median ICAPE value is $\sim$50\% higher at MesoDRY locations than MesoWET, reflecting a distribution closer to that for MCS environments (3f). 
Near--surface parcel CAPE values about MesoWET are higher (Fig.~S6c), but overall ICAPE better reflects the thermodynamic state across the full PBL \cite{Alfaro2015thermodynamic}, the lifting of which has been proposed as a dynamical control on MCS convection \cite{Alfaro2017low}. We find significantly higher thermodynamic instability prior to MCS cores than that originating from SM patches alone, indicating that favourable conditioning of the convective environment stems from the superposition of multiple mechanisms. Crucially though, such conditioning represents an enhancement of conditions about MesoDRY locations. Mesoscale wet patches meanwhile act as an inhibitor for mature MCSs, by suppressing favourable convective environments.

\begin{figure}[t]
    \centering
    \includegraphics[width=\textwidth]{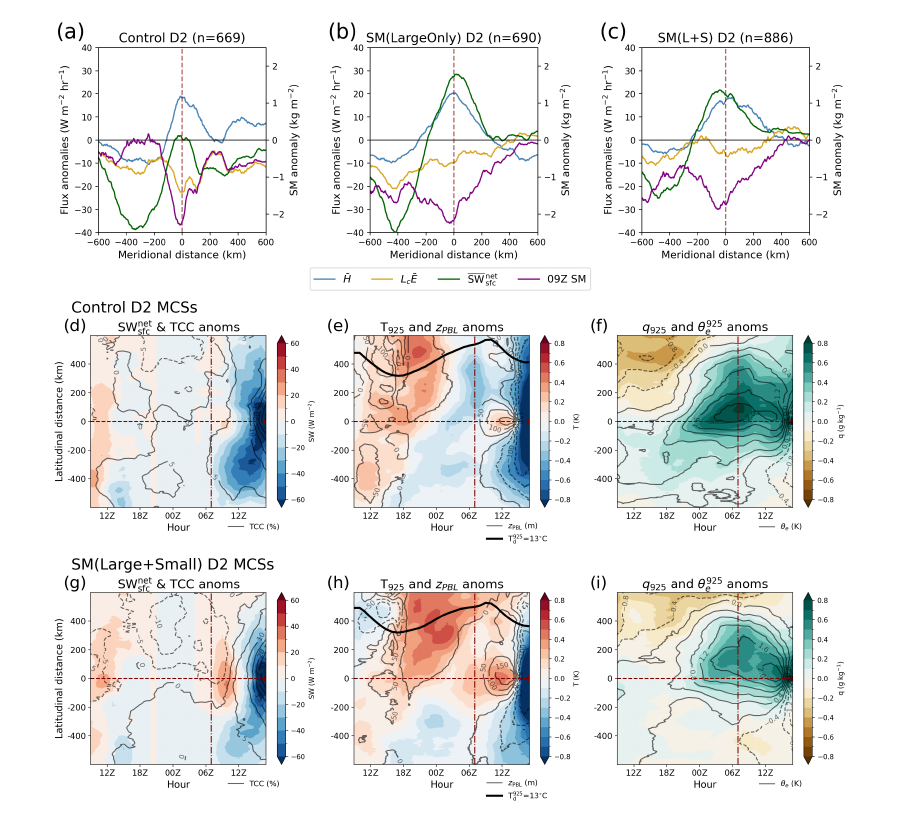}
    \caption{\textit{(a-c)} Zonal sections, for each simulation, of composite mean 09--12UTC flux anomalies at D2 locations of 17UTC Sahel MCS cores
    {; titles specify sample sizes}. \textit{(d-f)} Composite Hovmoellers of evolution prior to Control D2 core locations of anomalous \textit{(d)} shortwave radiation and total cloud cover (TCC); \textit{(e)} PBL height and 925hPa temperature; and \textit{(f)} 925hPa humidity and equivalent potential temperature ($\theta_e$). \textit{(g-i)} Repeated for SM(Large+Small) D2. 150km longitudinal slices used throughout
    {; vertical lines denote start of D2}.}
\end{figure}

As noted previously, we continue to find mature MCSs in the experiments following positive $H$ anomalies (Fig.~3a). Why then are the storm populations so different? Examining the sampled environments' spatial structure (Fig.~4a--c) shows that $H$ anomalies prior to MCS$_{\rm Exp}$ cores are driven by $>$200km peaks in surface net shortwave radiation ($\rm SW_{sfc}^{net}$), rather than dry soil patches. There are no colocated peaks in $L_cE$.

The importance of insolation in driving $H$ is confirmed by Hovmoellers (Fig.~4d--i) in advance of D2 MCS core occurrence. The SM(Large+Small) experiment (conditions closest to Control) shows a positive morning $\rm SW_{sfc}^{net}$ anomaly (
{Fig.~}4g) spanning $\sim$400km and located in a region of reduced cloud cover. Similar behaviour occurs on both days and in SM(LargeOnly) (
Fig.~S7
). Mesoscale shortwave variability is higher than in Control (Fig.~S1c), which shows weaker and smaller pre--storm anomalies (Fig.~4d), while cloud cover is increased. In all cases, $\rm SW_{sfc}^{net}$ plummets after 14UTC as MCS anvils develop and advance ahead of the convective cores.

Mesoscale positive PBL temperature and height anomalies are apparent before midday at all later core locations. These stem from the co--located $H$ anomaly, and generate convergence prior to storms (Fig.~1e). The primary difference in core--relative environments is low--level moisture and thermodynamic instability
{(4f,i)}, both of which increase prior to core occurrence. In agreement with observations \cite{Klein2020dry}, southerly flow is responsible for this build--up: for evening MCSs, monsoon flow the night before is the primary source of column moisture, which during the day reinforces PBL development from surface $H$ anomalies to provide an optimal convective environment. This mechanism is found in all simulations.

The primary role of dry mesoscale SM anomalies is thus to instigate the mechanism explored in Sec.~\ref{sec:patch_changes} by causing $H$ anomalies. When suppressed, this feedback can be driven by insolation from cloud--free slots. However, such patches are necessarily more diffuse and ephemeral than SM counterparts, reducing opportunities for the chain of PBL development to build conditions favourable for mature storms. This will reduce the MCS population, especially when combined with suppression of storm initiation from small--scale SM features (Fig.~2a). The absence of mesoscale SM variability also removes wet patches. MCSs which do establish in the experiments therefore also lose an inhibitor, instead benefiting from continued nocturnal monsoon flows and relatively uniform high low--level humidity. Their population therefore comprises fewer storms, but of comparable intensity to Control (Figs.~2b and~2c). Early morning relative intensification may then be due to modified land--surface interactions with cold pools \cite{Gentine2016role,Drager2020cold}, which Sahelian MCS populations are most sensitive to around dawn \cite{Maybee2025how}.


\section{Conclusions}
\label{sec:conclusions}

This study has investigated the sensitivity of MCSs to scales of SM heterogeneity, testing feedbacks observed over the Sahel \cite{Taylor2007observational,Klein2020dry} in MetUM simulations initialised from modified 06UTC SM fields. No other changes were made. In SM(LargeOnly) experiments where all sub--synoptic spatial SM variability was homogenised, we find a 23\% decrease in MCS numbers at the Day 1 diurnal peak of convection versus Control simulations. This is a result of suppressing mesoscale dry patches (MesoDRY), which provide favourable dynamic and thermodynamic conditions for mature storms. For SM(Large+Small) experiments where only mesoscale SM variability is suppressed, the decline in storm population is smaller (-13\%) due to higher rates of primary initiation than over homogenous SM, consistent with the role of small--scale SM gradients in enhancing MCS initiation \cite{Taylor2011frequency}. A full SM spectrum also contains mesoscale wet patches (MesoWET), which inhibit storms by suppressing favourable convective environments. These are also suppressed in both experiments, leading to comparable peak MCS intensities to Control despite declines in storm numbers.

Our results demonstrate the significant control SM variability can exert on MCS populations. As context, a similar experiment targeting the region's storms' sensitivity to cold pools \cite{Maybee2025how} demonstrated only a 4\% reduction in peak MCS population, but stronger changes to convective intensity. In the Sahel, SM heterogeneity therefore plays a stronger role in supporting MCS populations than cold pools, with the interplay of these sensitivities demanding future investigation. Significant influences of land--atmosphere interactions on MCS dynamics occur globally \cite{Barton2025storm}: our study highlights the relative magnitude of these effects.

In both experiments we find PBL development before MCS cores similar to that in Control prior to MCSs, and following dry SM patches. However, the crucial morning $H$ anomaly is instead driven by insolation over cloud--free slots. This has important implications for predictability. The Sahelian land--surface has a 2 week memory of MCS passage, with a lagged effect on rainfall up to 8 days \cite{Taylor2024multiday}, and can be utilised operationally for nowcasting \cite{Taylor2022nowcasting}. Feedbacks stemming from insolation lose this multi--scale predictability, thereby reducing the predictability of MCS behaviour. The land--atmosphere interactions in the Sahel are thus advantageous for local forecasting of convection versus regions with homogenous SM, and show the benefit of identifying and utilising these mechanisms for other regions with mesoscale surface heterogeneity.


\section*{Acknowledgements}

The research presented here was conducted within the NERC funded LMCS project (NE/ W001888/1). C.K. acknowledges funding from the NERC fellowship project COCOON (NE/X017419/1). We thank Doug Parker, Emma Barton, Brian Mapes and Yutong Lu for related discussions and comments
{, and two anonymous referees whose reviews have improved our paper}. All simulations were run on ARCHER2, the UK National Supercomputing Service \cite{Beckett2024archer2}, with data analysis conducted on JASMIN, the UK national collaborative data analysis facility.

\section*{Open Research}

Code and data underpinning this research are available publicly at \citeA{Maybee2025zenodo}.

\section*{Supporting Information}

Supporting Information can be found with the online version of this article.

\section*{Conflicts of Interest}

The authors have no conflicts of interest to disclose.

\bibliography{paper_draft}

\end{document}